\documentclass[submission, Chem]{SciPost}
\usepackage{silence}
\hypersetup{
    colorlinks,
    linkcolor={red!50!black},
    citecolor={blue!50!black},
    urlcolor={blue!80!black}
}

\usepackage[bitstream-charter]{mathdesign}
\DeclareSymbolFont{usualmathcal}{OMS}{cmsy}{m}{n}
\DeclareSymbolFontAlphabet{\mathcal}{usualmathcal}

\usepackage[english]{babel}
\usepackage[babel]{microtype}
\usepackage{adjustbox}

\graphicspath{{figures/}}

\addto\extrasenglish{%
    \def\equationautorefname~#1\null{Eq.~(#1)\null}%
}

\usepackage{siunitx}
\usepackage{nicefrac}
\usepackage[version=4]{mhchem}

\newcommand*{\order}[1]{\ensuremath{\mathcal{O}\left(#1\right)}}

\newcommand*{\an}[1]{\ensuremath{c_{#1}}}
\newcommand*{\cre}[1]{\ensuremath{c^\dagger_{#1}}}

\usepackage{mathtools}
\DeclarePairedDelimiter{\ket}{\lvert}{\rangle}

\begin{document}

\begin{center}{\Large \textbf{
The seniority quantum number in Tensor Network States
}}\end{center}

\begin{center}
K. Gunst\textsuperscript{1$\star$},
D. Van Neck\textsuperscript{1},
P.A. Limacher\textsuperscript{2} and
S. De Baerdemacker\textsuperscript{3}
\end{center}

\begin{center}
{\bf 1} Ghent University, Ghent, Belgium
\\
{\bf 2} SAP Security Research, Karlsruhe, Germany
\\
{\bf 3} Department of Chemistry, University of New Brunswick, Fredericton, Canada
\\
${}^\star$ {\small \sf klaasgunst@gmail.com}
\end{center}

\begin{center}
\today
\end{center}


\section*{Abstract}
{\bf
  We employ tensor network methods for the study of the seniority quantum
  number -- defined as the number of unpaired electrons in a many-body wave
  function -- in molecular systems. Seniority-zero methods recently emerged as
  promising candidates to treat strong static correlations in molecular
  systems, but are prone to deficiencies related to dynamical correlation and
  dispersion.
  We systematically resolve these deficiencies by increasing the allowed
  seniority number using tensor network methods. In particular, we investigate
  the number of unpaired electrons needed to correctly describe the binding of
  the neon and nitrogen dimer and the $\mathbf{D_{6h}}$ symmetry of benzene.
}

\vspace{10pt}
\noindent\rule{\textwidth}{1pt}
\tableofcontents\thispagestyle{fancy}
\noindent\rule{\textwidth}{1pt}
\vspace{10pt}

\section{Introduction}
The quantum mechanical characterization of molecular systems is highly
nontrivial due to its many-body character. The need for approximate
methods arises for all but the smallest problems.  When choosing a suitable
approximate method, a consideration has to be made between the accuracy and the
complexity of the method. The well-known Hartree-Fock (HF) method provides a
mean-field solution for molecular systems rather cheaply. The deficit between
the exact ground state energy and the approximate Hartree-Fock energy is an
important quantity in quantum chemistry and is called the correlation energy.
Often, the distinction between strong (static and nondynamical) and weak
(dynamical) correlation is made~\cite{Helgaker_2014_ch5, Bartlett_2007,
Bartlett_2007b, Sinano_lu_1963}.  Although both strong and weak correlations
are electronic by nature, they have a different origin; the latter 
results mainly from the dynamical short-range correlations of electrons,
whereas static correlation originates from near-degeneracies of several
rivaling electron configurations. Many approximate methods often only excel
in capturing one type of correlation. For example, the complete active space
self consistent field method (CASSCF)~\cite{Roos_1980, Roos_2009} and density
matrix renormalization group (DMRG)~\cite{White_1992, White_1993,White_1999}
are capable of capturing strong correlations within a chosen active space,
while coupled cluster (CC)~\cite{Bartlett_2007, Cizek_1966} and perturbative
methods~\cite{Moller_1934} are more suitable for dynamical correlations. In an
effort to capture both types of correlation, combinations of these methods have
also been developed such as CASPT2 (CAS with perturbation theory up to second
order)~\cite{Andersson_1990, Andersson_1992}, DMRG-CASPT2~\cite{Kurashige_2011,
Wouters_2016}, DMRG-NEVPT2 (DMRG with second-order N-electron valence state
perturbation theory)~\cite{Sharma_2014}, p-DMRG~\cite{Guo_2018}, MRCC
(multireference coupled cluster)~\cite{Musial_2011, Kohn_2012,
Evangelista_2018} and DMRG-TCC (DMRG-tailored coupled
cluster)~\cite{Hino_2006, Veis_2016, Faulstich_2019}.

The majority of contemporary electronic structure methods start from a
reference state, typically the single-reference HF ground state, and
systematically build in correlations by considering elementary excitations from
this reference.  The conventional approach is to consider particle-hole (ph)
excitations from the HF ground state, as is common in CC~\cite{Bartlett_2007} or
truncated configuration interaction (CI) methods~\cite{Helgaker_2014_ch5}.
This way, it is possible to construct a hierarchy of multiple $n$-ph
excitations which are assumed to be decreasing in importance with increasing
$n$. Although tailor-made for dynamical correlations, e.g.\ in CC 
theory, it is impractical for static (or non-dynamical) correlation.  It was
recently observed~\cite{Bytautas_2011} that the \emph{seniority scheme} is much
better suited to capture static correlations associated with the entanglement
structure of single-bond breaking processes. Defined as the number of unpaired
electrons in a Slater determinant, the seniority quantum number organizes the
Hilbert space by the amount of broken closed-shell singlet pairs with respect
to a set of (doubly degenerate) spin orbitals.  For molecular systems dominated
by singlet-pairs bond structures, it was
shown that most of the strong static correlation in a system can already be
captured in the subspace spanned by all determinants with zero seniority (no
unpaired electrons)~\cite{Weinhold_1967, Weinhold_1967b, Bytautas_2011,
Kutzelnigg_2012, Limacher_2013, Alcoba_2014, Limacher_2016}.  Although this
tremendously reduces the dimension of the Hilbert space at hand, finding the
exact doubly occupied configuration interaction (DOCI) wave function is still
an exponentially scaling problem.  At first glance, the seniority scheme seems
only marginally more manageable than the full problem. Interestingly, the
antisymmetric product of one-reference orbital geminals
(AP1roG)~\cite{Limacher_2013,Boguslawski_2014a, Boguslawski_2014b,
Boguslawski_2014c}, also known as pair-coupled cluster doubles
(pCCD)~\cite{Stein_2014, Henderson_2014, Shepherd_2016, Henderson_2015},
appears to provide a reliable approximation to the DOCI ground state energy
for a wide range of molecular systems~\cite{Shepherd_2016} while staying
computationally tractable at a mean-field scaling computational
cost~\cite{Limacher_2013, tecmer_2015}.

Notwithstanding its salient features, there remain several challenges that need
to be overcome in order to make the AP1roG wave function quantitatively
accurate.  The outstanding challenges, which are shared by all methods
expressed in the seniority scheme, are (i) the incorporation of
dynamical correlation and (ii) the choice of a preferential orbital set, also
referred to as the orbital optimization (OO) problem.   Another challenge (iii)
is the apparent lack of London dispersion correlations in the seniority-zero
methods which are crucial to model large-size molecular systems.  

The lack of dynamical correlation in the zero seniority wave functions is well
illustrated by the poor description of the correlation energy of the \ce{Ne}
atom, as well as the near-constant parallelity error in the bond dissociation
curve of the nitrogen dimer~\cite{Limacher_2013,Limacher_2014a,mok_1996}.
Dynamical correlation is generally encoded in a set of Slater
Determinants with a few ph excitations from the HF reference state.
Consequently, methods targeting these excitations, such as (multi-reference)
perturbation theory~\cite{Limacher_2014a,kobayashi_2010}, linearized
CC~\cite{Henderson_2014,boguslawski_2015},  extended random phase
approximiation (ERPA)~\cite{pastorczak_2015} or selected configuration
interaction (CI)~\cite{vanraemdonck_2015}, are very well equipped to capture
those correlations.  However, systematic generalizations of these methods in
order to include dynamical correlation from higher-order ph excitations prove
either technically and computationally demanding, or break the size consistency
of the reference AP1roG wave function. 

The concept of paired orbitals is dependent on the choice of the basis
orbitals~\cite{Alcoba_2013}.  Current optimization methods unfortunately result
into a single unique set of spin orbitals, which can lead to nonphysical
symmetry breaking effects in resonating bond structures, such as the aromatic
structures in benzene \cite{Boguslawski_2014c}, or incorrect characterizations
of covalent triplet-bond couplings, such as in the nitrogen
dimer~\cite{limacher_2014b}.  

Regardless the correct description of the static
correlations associated with bond-dissociation processes, seniority-zero methods
have recently been identified as essentially free from London dispersion energy
\cite{brzek_2019}, which is remarkable given that 2-electron systems are
exactly described by (orbital optimized) seniority-zero methods, capturing the
non-covalent Lennard-Jones ${1}/{R^6}$ behavior of the dispersion energy in the
large $R \to \infty$ separation limit of the hydrogen dimer.

In order to obtain a global understanding of the deficiencies of the
seniority-zero methods, it is quintessential to include all possible
broken-pair excitations from higher seniority sectors in a
systematic way.  Higher seniority subspaces have been studied in the past years
using CI approaches, \cite{Bytautas_2011, Wahlen_Strothman_2017, Bytautas_2015,
Stein_2014, Alcoba_2014}, or energy renormalization group (ERG) approaches
\cite{Bytautas_2018}.  The limiting factor of these methods is the pernicious
computational scaling whenever no truncation in the Slater determinants is
considered.  While dynamical correlation is typically included with just a few
ph excitations from the HF reference state, corresponding to low-seniority
quantum states, it is not clear at present how many broken pairs are needed to
restore the correct symmetries or include London dispersion.  As a result,
there is a need for an analytic method that can assess seniority
non-zero contributions in a systematic way at a favorable computational
scaling.  

In this paper, we use the concept of seniority in junction with tensor
network states.  In contrast to many other quantum many-body methods, tensor
network states consider the whole collection of Slater determinants, and
approximate the exact quantum states by restricting the amount of entanglement
between local degrees of freedom. Tensor network states are capable of encoding
local symmetries of quantum states~\cite{orus_2019}; therefore they provide
a good framework to investigate broken pair excitations, as seniority can be
related to the irrep label of the $su(2)$ quasi-spin algebra \cite{talmi_1993}.
In  practice, the idea is to perform DMRG in a subspace of the Hilbert space up
to a fixed global seniority quantum number, and increase the seniority quantum
number until full convergence of the correlation energy is obtained.  This
procedure will be explained in detail in \autoref{sec:method}.  In the
proceeding sections, we will present results for the nitrogen dimer
(\autoref{sec:Nitrogen}), benzene (\autoref{sec:benzene}) and the neon dimer
(\autoref{sec:Neon}), to discuss higher-seniority properties of dynamical
correlation, symmetry breaking/restoration and dispersion respectively.   

\section{Methodology}
\label{sec:method}
\subsection{Tensor networks}
\label{sec:TNS}
Pioneered by Steve White in 1992~\cite{White_1992, White_1993}, tensor networks
have proven to be a natural language for the entanglement in strongly
correlated many body systems. In the tensor network state, each tensor
represents a `local' physical degree of freedom. By connecting them in a
network, correlations between the different physical degrees of freedom can be
encoded through their virtual degrees of freedom. The exact layout of the
network influences the entanglement structure that can be represented; it is
easier to correlate physical degrees of freedom that are close in the network. 

First and foremost, these tensor network methods have established themselves in
the field of condensed matter physics as a wide range of successful tensor
networks have been developed for numerous problems. Some notable examples are
matrix product states (MPS)~\cite{White_1992, White_1993, Ostlund_1995,
Dukelsky_1998, Verstraete_2004b}, projected entangled pair states
(PEPS)~\cite{Verstraete_2004} and the multiscale entanglement renormalization
ansatz (MERA)~\cite{Vidal_2007}.  They all provide, in their own way, an
efficient representation of certain entanglement structures. 

In quantum chemistry, tensor networks have also proven their worth in the study
of molecules with strong correlations~\cite{White_1999, Chan_2002, Legeza_2003,
Szalay_2015, Sharma_2012, Wouters_2014b, Wouters_2014a, Brabec_2020, Murg_2010,
Murg_2015, Nakatani_2013}. Quantum chemists don't traditionally study molecules
in a Hilbert space built from completely local basis functions (e.g.\ a grid in
three dimensions), but atomic orbital basis sets such as Gaussian-type or
Slater-type orbitals are used.  These sets give electrons the right flexibility
needed for chemistry while the basis size is kept small.  On the flip side, the
loss of locality in the basis functions makes a suitable network for the
entanglement between the physical degrees of freedom less straightforward than
for most condensed matter problems.  Furthermore, in an atomic orbital basis
set, the long range two-body coulomb interactions in the Hamiltonian become
four-point interactions. The loss of locality and the need
for an efficient evaluation of the Hamiltonian has ensured that the most simple
networks are still the most preferred ones. The density matrix renormalization
group (DMRG) is, by far, the most popular tensor network method in quantum
chemistry and corresponds with the optimization of the linear MPS.  Another
option for a simple tensor network is the three-legged tree tensor network
state (T3NS)~\cite{Gunst_2018, Gunst_2019}. It is a subclass of the more general
tree tensor networks (TTNS)~\cite{Murg_2010, Murg_2015, Nakatani_2013} and was
recently introduced by some of us. In this paper, we use these two networks for
the study of several chemical systems in restricted seniority subspaces. In the
next sections we explain the implementation of restricted seniority for the
case of DMRG. However, the ideas are readily adaptable to T3NS and were
implemented for both cases in our in-house T3NS-code~\cite{T3NSsourcecode}.

\subsection{Seniority and tensor networks}
The non relativistic quantum chemical Hamiltonian to study  is given by
\begin{equation}
  H = \sum_{ij} t_{ij} \sum_{\sigma} \cre{i\sigma} \an{j\sigma} + 
  \frac{1}{2} \sum_{ijkl}V_{ijkl} \sum_{\sigma\tau}\cre{i\sigma}\cre{j\tau} \an{l\tau}\an{k\sigma}~,
  \label{eq:HQchem}
\end{equation}
where $i, j, k$ and $l$ are the indices of the orbitals and $\sigma$ and $\tau$
index the spin of the electrons. This Hamiltonian showcases several symmetries,
e.g.\ the particle conservation and total spin symmetry of the electrons. These
symmetries can be easily exploited in tensor networks by writing the different
tensors in an invariant form under group action of the symmetry~\cite{
McCulloch_2007, McCulloch_2002, Singh_2010b, Singh_2010a, Wouters_2014b,
Weichselbaum_2012, Singh_2012, Hubig_2018, Toth_2008, Gunst_2019}. Although the
seniority is not a symmetry of the quantum chemical Hamiltonian, it is still
possible to apply the same idea. In this case, we write each tensor in the
network in an invariant form for the seniority. For example, the tensors of
rank three present in the MPS can be made invariant by imposing the following
restriction for the tensor elements:
\begin{align}
  T_{a,b,c} &= 0, & \mathrm{if}\ \nu_a + \nu_b &\neq \nu_c~,
  \label{eq:seniorityrestriction}
\intertext{or graphically}
  \adjustimage{valign=c}{MPS_seniority1tens} &= 0, & \mathrm{if}\ \nu_a + \nu_b &\neq \nu_c~,
  \label{eq:seniorityrestriction_graph}
\end{align}
where $a$, $b$ and $c$ denote the (physical or virtual) degrees of freedom of
$T$ and $\nu_a$, $\nu_b$ and $\nu_c$ are their respective seniority numbers
which is well-defined; each state in the degrees of freedom $a$, $b$, and $c$
are eigenstates of the seniority operator. In this example, the seniority
number of the first two degrees of freedom of the tensor sum up to the
seniority number of the third one. This reflects the fact that seniority is an
additive feature, as unpaired electrons from different orbitals all contribute
to the total seniority of the state.  It is clear that this restriction implies
a kind of \emph{flow} for the seniority number in the network which is
indicated in \autoref{eq:seniorityrestriction_graph} by the directed edges. An
example of an MPS wave function built from three of these invariant tensors
with the flow indicated is given by
\begin{align}
  \ket{\Psi} &= \sum_{\alpha,\beta,a,b,c,f} A_{\text{vac},a,\alpha}
  B_{\alpha, b, \beta} C_{\beta, c, f} \ket{abc}
  \label{eq:MPSseniority_alg} \\
             &= \sum_{a,b,c,f} \adjustimage{valign=c}{MPS_seniority2}~\ket{abc}
  \label{eq:MPSseniority}\\
             &= \sum_{f} \ket{\phi_f}
  \label{eq:MPSseniority_split}
  \intertext{where}
  \ket{\phi_f} &= \sum_{a,b,c}
  \adjustimage{valign=c}{MPS_seniority2}~\ket{abc}~.
  \label{eq:MPSseniority_splitval}
\end{align}
The physical degrees of freedom (the occupancies of the spatial orbitals) are
denoted by $a$, $b$ and $c$ in this example and have a seniority $\nu \in \{0,
1\}$. $\alpha$ and $\beta$ are virtual degrees of freedom. The vacuum state
enters the MPS at the leftmost degree of freedom (vac) and has a seniority $\nu
= 0$. The final degree of freedom $f$ represents different parts of the total
wave function $\ket{\Psi}$.  The restriction on the tensors given in
\autoref{eq:seniorityrestriction_graph} ensures that each part $\ket{\phi_f}$
given by \autoref{eq:MPSseniority_splitval} has a well-defined seniority
number, i.e.  each Slater determinant $\ket{abc}$ with a non-zero contribution
for a particular $\ket{\phi_f}$ has the same seniority number. We
can also easily ensure that each $\ket{\phi_f}$ has a unique seniority number by
summing any $\ket{\phi_f}$ states with matching seniority numbers. This results
in an orthogonal set of $\ket{\phi_f}$ (but not orthonormal). The graphical
depiction implies for each connected edge a summation over its corresponding
indices; the summation over $\alpha$ and $\beta$ are implied in
\autoref{eq:MPSseniority}. This graphical notation is widely used in the
tensor network language~\cite{Verstraete_2004, Schollwock_2011b, Orus_2014,
Biamonte_2017, Gunst_2019, orus_2019}.

The only difference with implementing a $\operatorname{U}(1)$-symmetry of the
system, e.g.\ particle conservation or conservation of the spin projection, is
the needed summation over the states of the final edge $f$ in
\autoref{eq:MPSseniority}. This is necessary as the seniority is \emph{not} a
conserved quantum number.  Eigenstates of the Hamiltonian are not necessarily
eigenstates of the seniority operator and the target state can be a linear
combination of Slater determinants with different seniority numbers. To target
such a state, the final states at edge $f$ are a set of eigenstates of the
seniority operator which combine to the targeted state when summed. In
contrast, for a conserved quantity of the system such as the particle
conservation, there is only one state $\ket{\phi_{f}}$ needed.

The set of possible seniority numbers for the wave function is
\begin{equation}
  \Omega = \left\{n \in \mathbb{N}:
    \begin{matrix}
      n \bmod 2 = N_\mathrm{tot} \bmod 2\\
      \left|N_\uparrow - N_\downarrow\right| \leq n \leq \min\left(N_\mathrm{tot}, 2 k - N_\mathrm{tot}\right)
    \end{matrix}
  \right\}~,
\end{equation}
with $k$ the number of spatial orbitals, $N_\uparrow$ $(N_\downarrow)$ the
number of electrons with spin up (down) and $N_\mathrm{tot}$ the total number
of electrons.  For every renormalized state in the last edge, we have $\nu_f
\in \Omega$.  By restricting  $\nu_f$ to a subset $S$, i.e.\ $\nu_f \in S
\subseteq \Omega$, ground states in seniority-restricted subspaces can be
targeted. The weight of each seniority subspace for the total wave function can
be readily calculated as $|c_{\nu_f}|^2 = \langle \phi_{f} \vert
\phi_{f}\rangle$.

In a similar fashion, one could also use other non-conserved quantum
numbers than the seniority. For example we could use the excitation number with
respect to the Hartree Fock wave function.  By only allowing Slater
determinants with a certain amount of excitations, tensor networks can be used
as an approximate configuration interaction (CI) solver with arbitrary allowed
excitation levels.

\subsubsection{Suboptimal decomposition}
\label{sec:subopt}
When using a wave function ansatz as shown in \autoref{eq:MPSseniority}, we
impose a restriction on the left renormalized states at each splitting of the
network. Due to the fact that a vacuum state enters in the left most edge
(vac) and all tensors used are invariant under the seniority operator,
the left renormalized states need to have a well defined seniority number. This
restriction does not hold for the right renormalized states since multiple
states with different seniority exit at the right most edge $f$.

This restriction results in the need of a possibly larger bond dimension than
when discarding seniority. We illustrate this using a wave function with three
electrons in three spatial orbitals:
\begin{align}
  \label{eq:3site_seniority}
  \ket{\Psi} &= \frac{1}{\sqrt{2}}\left[\ket{\uparrow, \downarrow} \otimes \ket{\uparrow}
  + \ket{-, \uparrow\downarrow} \otimes \ket{\uparrow}\right]\\
  \label{eq:3site_schmidt}
             &= \frac{1}{\sqrt{2}}\left[\ket{\uparrow, \downarrow} +
             \ket{-, \uparrow\downarrow}\right] \otimes \ket{\uparrow}~.
\end{align}
In \autoref{eq:3site_schmidt}, the Schmidt decomposition for a partitioning
between the first two and the last orbital is given. At this partitioning only
a virtual bond dimension of one is needed to represent the state. However, when
we impose that the left states, i.e.\ the states in the first two orbitals, of
the decomposition should also be eigenstates of the seniority operator, the
needed bond dimension at this partitioning increases to two, confer
\autoref{eq:3site_seniority}.

\subsection{DOCI and tensor networks}
Restricting the calculation to configurations with $\nu = 0$, i.e.\ all
electrons are paired, is easily done with the aforementioned method. However,
it is more efficient to directly implement the quantum chemical
Hamiltonian projected on the DOCI-subspace where only paired electrons are
allowed. The DOCI-Hamiltonian is given by
\begin{equation}
  {H}_\mathrm{DOCI} = 2\sum_{i} t_{ii} {n}_i +
  \sum_{ij} \left(2V_{ijij} - V_{ijji}\right) {n}_i {n}_j +
  \sum_{i \neq j} V_{iijj} {b}^\dagger_i {b}_j~,
  \label{eq:HDOCI}
\end{equation}
where ${b}^\dagger_i$ and ${b}_i$ are the bosonic pair creation and
annihilation operators and ${n}_i$ is the pair number operator at orbital
$i$. They are given by
\begin{align}
  {b}^\dagger_i &= \cre{i\uparrow}\cre{i\downarrow}~, &
  {b}_i &= \an{i\downarrow}\an{i\uparrow}
\end{align}
and
\begin{equation}
  {n}_i = {b}^\dagger_i{b}_i~.
\end{equation}
This Hamiltonian only scales quadratically with the number of orbitals in
contrast with the quartic scaling of the full Hamiltonian. TNS calculations in
the DOCI subspace can be performed with a lower polynomial scaling, as stated in
\autoref{tab:DOCIscaling} for the DMRG and T3NS.

\begin{table}[!ht] 
  \centering 
  \begin{tabular}{r c c} & CPU time & Memory\\
    \cline{2-3} & &
    \vspace{-0.2cm}
    \\
    QC-DMRG: & $\order{k^4D^2 + k^3D^3}$ & $\order{k^3D^2}$\\
    DOCI-DMRG: & $\order{k^2D^3}$ & $\order{k^2D^2}$
    \vspace{0.2cm} \\
    QC-T3NS: & $\order{k^4D^2 + k^3D^4}$ & $\order{k^3D^2 + kD^3}$\\
    DOCI-T3NS: & $\order{k^2D^4}$ & $\order{k^2D^2 + kD^3}$\\
  \end{tabular}
  \caption{%
    Resource requirements for DMRG and T3NS with renormalized operators for the
    full quantum chemical Hamiltonian (\autoref{eq:HQchem}) or the DOCI
    Hamiltonian (\autoref{eq:HDOCI}) without orbital optimization for $k$
    spatial orbitals. The maximal virtual bond dimension is denoted by $D$.
  } 
  \label{tab:DOCIscaling}
\end{table}

We find that DOCI ground state wave functions have in general lower
entanglement than their corresponding full configuration interaction (FCI)
ground state wave function; accurate results for DOCI can be obtained with a
much lower bond dimension. The synergy between the lower polynomial scaling and
the lower bond dimension needed, makes DOCI-TNS very fast and a good option for
initializing tensor network calculations in the FCI space. For example, DOCI
calculations without orbital optimization with 162 electron pairs and 261
spatial orbitals can be executed in a several minutes on a common
laptop~\cite{DOCIcalc}.

Not only DOCI-TNS but also general seniority-restricted tensor network
calculations can provide interesting approximations from a computational
viewpoint, although they have the same polynomial scaling as unrestricted
calculations.  They can converge faster than the latter due to the lower
entanglement present in the wave function; hence a lower bond
dimension is needed. This is generally the case for $\nu \leq 2$ and (to a lesser
extent) $\nu \leq 4$ calculations.\footnote{Bear in mind that, while these
calculations converge faster, they do not converge to the FCI limit.}

In contrast, using high seniorities will result in a loss of efficiency when
compared to seniority-unrestricted tensor network calculations due to the extra
bookkeeping needed and the suboptimal Schmidt decomposition (see
\autoref{sec:subopt}).  Especially the latter is detrimental as the
sought-after wave function at high seniority calculations is similarly
entangled as the exact solution, but the tensor network ansatz can not capture
it as efficiently as the equivalent unrestricted ansatz.  As
such, these types of calculations primarily provide a means to analyze the
need for broken pairs in chemical systems. They should not necessarily be
viewed as an efficient way to approximate the FCI solution. The presented
tensor network method allows to investigate the number of broken pairs needed
for recovering a qualitative correct picture when one attempts to correct
existing seniority-zero methods.

\section{Applications}
\label{sec:applications}

We discuss some calculations with the seniority-restricted tensor network code.
As these calculations are orbital dependent, several types of orbitals are
considered. The effect of allowing progressively more broken pairs is also
studied within each orbital set. In \autoref{sec:Nitrogen} and
\autoref{sec:Neon}, the dissociation of the nitrogen and neon dimer are
considered, respectively.  \autoref{sec:benzene} discusses the benzene
molecule, a system demonstrating artificial $D_{6h}$ symmetry breaking in the
seniority-zero subspace~\cite{Boguslawski_2014c}.

Coupled cluster natural orbitals and Löwdin orthogonalized atomic orbitals
are obtained with PySCF~\cite{Sun_2020, Sun_2017, Sun_2015}. DOCI-optimized
orbitals are generated through an in-house DOSCF code and were carefully
checked to correspond to the lowest possible DOCI energy, i.e.\ the global
minimum~\cite{limacher_2014b}.  The  seniority-restricted tensor network
calculations were executed with our own T3NS-code~\cite{T3NSsourcecode}.  All
seniority-restricted tensor network calculations are MPS calculations. We
exploit the spin symmetry and the reported bond dimensions for the tensor
networks are \emph{reduced} bond dimensions; renormalized states belonging to
the same multiplet are represented by one reduced renormalized state, thus
reducing the needed bond dimension.  Seniority-restricted tensor network
calculations are, just as regular tensor network calculations, not exact; the
accuracy can be controlled by the bond dimension. The following calculations
use a large enough bond dimension to ensure quantitatively accurate potential
energy surfaces.

\subsection{Nitrogen dimer}
\label{sec:Nitrogen}

\begin{figure}[!ht]
  \centering
  \includegraphics{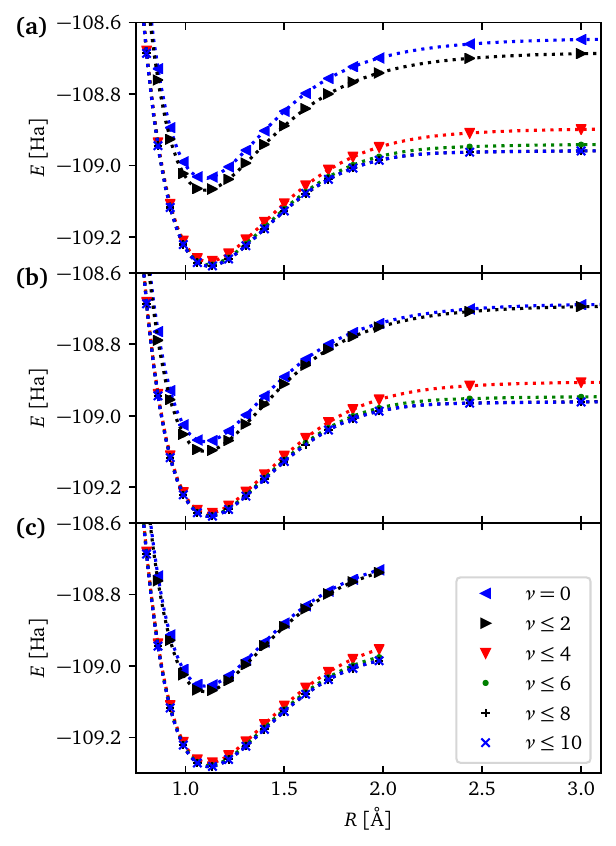}
  \caption{%
    Dissociation curves at different seniority subspaces for the nitrogen
    dimer. Results for canonical orbitals (a), DOCI-optimized orbitals (b) and
    CCSD natural orbitals (c) are given. For (c), only results where CCSD
    converged are plotted.
  }
  \label{fig:N2}
\end{figure}

Characterized by a triple bond breaking, the nitrogen dimer is a much visited
test case for new quantum chemical methods, and has already been investigated
as such in the seniority framework by Bytautas \emph{et al.}~\cite{Bytautas_2011} 
using an active space in the cc-pVDZ basis with $D_{2h}$-symmetry adapted MOs.
Here, we study the nitrogen dimer in a cc-pVDZ basis set with all electrons
correlated, however the DMRG results are qualitatively similar to the results
in \cite{Bytautas_2011}.  Seniority-restricted spin-adapted DMRG with a reduced
bond dimension up to a 1000 is used to optimize the ground state in the
different subspaces. The allowed seniority increases from 0 (DOCI) up to 10 for
the largest calculations, allowing 5 electron pairs to be broken. In
\autoref{fig:N2}, the dissociation curves are given for calculations within the
different seniority subspaces. Calculations were performed for canonical
orbitals (\autoref{fig:N2}a), DOCI-optimized orbitals (\autoref{fig:N2}b) and
CCSD natural orbitals (\autoref{fig:N2}c).  Although the DOCI-optimized
orbitals are optimized for the seniority-zero subspace specifically they also
perform better in higher seniority subspaces, albeit marginally. Eventually for
$\nu \leq 8$ and onward, all orbital sets give quasi-FCI energies. 

\begin{figure}[!ht]
  \centering
  \includegraphics{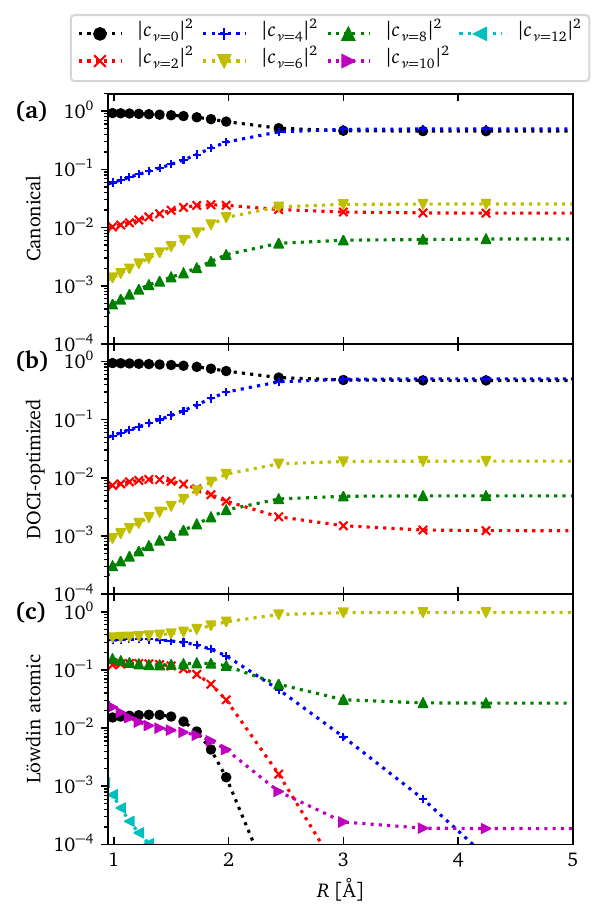}
  \caption{%
    Weights of the different seniority subspaces for the ground state wave
    function of the nitrogen dimer.  Results for canonical orbitals (a),
    DOCI-optimized orbitals (b) and Löwdin orthogonalized atomic orbitals (c)
    are given.
  }
  \label{fig:N2_ci}
\end{figure}

In Ref.~\cite{Boguslawski_2014b}, it is shown that the seniority-two sector
decouples from the seniority-two-plus-zero sector up to first order for
DOCI-optimized orbitals; only a small correction should occur due to the
introduction of single broken pairs in this orbital set. Putting this first
order decoupling to the test, we notice indeed a small energy correction for
the DOCI-optimized orbitals, smaller than for canonical orbitals. In
\autoref{fig:N2_ci}, the weights of the different seniority subspaces are
plotted for the ground state in both canonical and DOCI-optimized orbitals.
It is yet another illustration that for DOCI-optimized orbitals
(\autoref{fig:N2_ci}b) the seniority-two subspace is less important than for
canonical orbitals (\autoref{fig:N2_ci}a).
However, a first order decoupling is not an exact one; there are other orbital
sets possible which give even smaller energy corrections. This is illustrated
by the natural orbitals (\autoref{fig:N2}c) which give even smaller energy
corrections when allowing single broken pairs in this system.
 
As a last observation we note that the largest change in energy occurs
when including the seniority-four subspace, and this for all orbital sets in
\autoref{fig:N2}. This trend was also noticed in Ref.~\cite{Bytautas_2011} for
the nitrogen dimer in nonlocal orbitals. When including up to seniority four
the energies are close to converged around the binding distance for increasing
seniority numbers; however, the binding energy itself is still overestimated
due to missing dynamical correlation at the dissociation (values are given in
\autoref{tab:N2_dissoc} for both canonical and DOCI-optimized orbitals).

\begin{table}[ht] 
  \centering 
  \begin{tabular}{r | l l l l l l}
    & $\nu = 0$ & $\nu \leq 2$ & $\nu \leq 4$ & $\nu \leq 6$ & $\nu \leq 8$ & $\nu \leq 10$\\
    \cline{2-7}
    canonical & \num{424} & \num{382} & \num{377} & \num{338} & \num{322} & \num{322}\\
    DOCI-optimized & \num{383} & \num{405} & \num{367} & \num{333} & \num{321} & \num{321}\\
  \end{tabular}

  \caption{%
    Binding energies in \si{\milli\hartree} for seniority-restricted
    calculations in both canonical and DOCI-optimized orbitals.
  } 
  \label{tab:N2_dissoc}
\end{table}

Intuitively, we would expect a much larger error when excluding the
seniority-six subspace as Hund's rule dictates dissociation to two nitrogen
atoms with each three unpaired electrons.  However, seniority and pairing is an
orbital-dependent concept~\cite{limacher_2014b}; we need to keep in mind that
Hund's rule applies to a nitrogen atom with orbitals localized around that
atom.  To study the interpretation of Hund's rule in non-local orbitals, we
consider a toy model of two sets of three orbitals $(p_x, p_y, p_z)$ and
$(p'_x, p'_y, p'_z)$. Each set of orbitals mimics the local $p$-orbitals of
each nitrogen atom which are singly occupied and couple together to a
$S=\nicefrac{3}{2}$ state, as dictated by Hund's rule. Our tensor network
calculations target over the whole dissociation curve a singlet state for the
dimer, so the two toy-nitrogen atoms should couple as $\left[\nicefrac{3}{2},
\nicefrac{3}{2}\right]^0$.  Mimicking non-local orbitals, we rotate the orbitals
pairwise as follows:
\begin{align*}
  \pi_1 &= p_x \cos\theta + p'_x \sin\theta,\qquad
  \pi^*_1 = -p_x \sin\theta + p'_x \cos\theta\\
  \pi_2 &= p_y \cos\theta + p'_y \sin\theta,\qquad
  \pi^*_2 = -p_y \sin\theta + p'_y \cos\theta\\
  \sigma &= p_z \cos\theta + p'_z \sin\theta,\qquad
  \sigma^* = -p_z \sin\theta + p'_z \cos\theta~.
\end{align*}
\begin{figure}[!ht]
  \centering
  \includegraphics{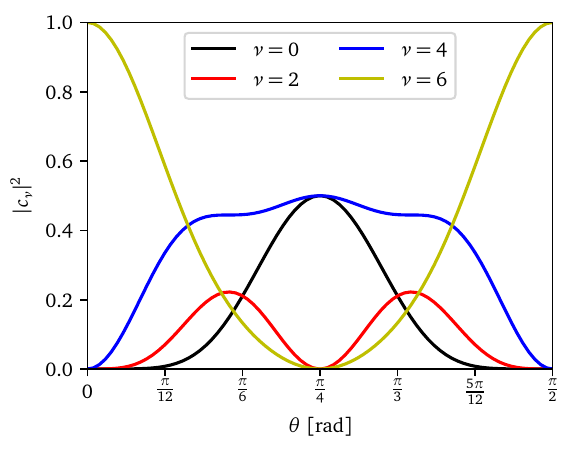}
  \caption{\label{fig:Hund}Toy model of two nitrogen atoms with both $S =
    \nicefrac{3}{2}$ respecting Hund's rule. The two atoms couple together to a
    singlet. The figure represents the weights of the different seniority
    sectors for local $\left(0, \nicefrac{\pi}{2}\right)$, delocalized
    $\left(\nicefrac{\pi}{4}\right)$ orbitals and everything in between.
  }
\end{figure}

In \autoref{fig:Hund}, the weights of the different seniority sectors is given
for the $\left[\nicefrac{3}{2}, \nicefrac{3}{2}\right]^0$ coupled toy wave function in
function of the rotation angle $\theta$. As can be seen in this model
seniority-six is actually of no importance when working with delocalized
orbitals $\left(\theta = \nicefrac{\pi}{4}\right)$.  Instead, the correct
dissociation can be described with only seniority-zero-plus-four and both
seniorities equally important.

Both canonical orbitals and the DOCI-optimized orbitals are delocalized for the
$2p$-orbitals in this system. This dominating importance of the seniority zero
and four for the wave function at dissociation is very clear in
\autoref{fig:N2_ci}a and \autoref{fig:N2_ci}b. The other seniority sectors have
very small contributions in comparison. As an illustration, we also included
calculations with Löwdin orthogonalized atomic orbitals in
\autoref{fig:N2_ci}c. As these orbitals are localized, it corresponds with
$\theta = 0$ in \autoref{fig:Hund}. These orbitals do give rise to a very
important seniority-six subspace at dissociation, as predicted by Hund's rule.
Evenmore, all seniority sectors smaller than six express a superexponential
decay.

\subsection{Benzene}
\label{sec:benzene}
\begin{figure}[!ht]
  \centering
  \includegraphics{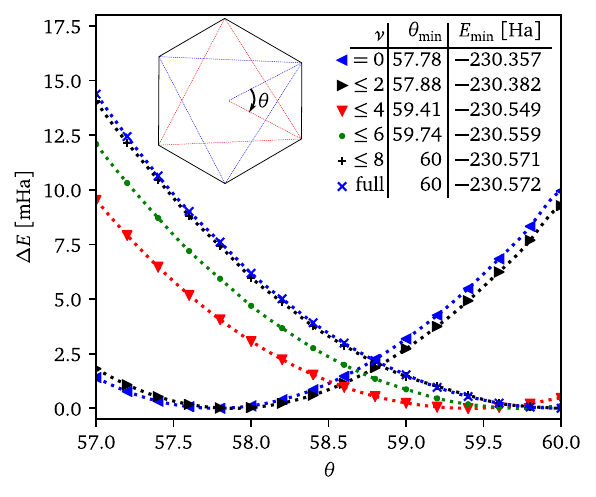}
  \caption{%
    Benzene in a STO-6G basis set for different distortion angles. The minimal
    energies and the corresponding distortion angles for increasing seniority
    subspaces are given in the inset. A graphical depiction of the in-plane
    distortion of the aromatic ring in benzene is also shown.
  }
\label{fig:benzene}
\end{figure}

In this section, the in-plane distortion of benzene is studied. The exact nature
of the distortion is given in the inset in \autoref{fig:benzene} and is
characterized by the angle $\theta$. At $\theta = \ang{60}$ the $D_{6h}$
symmetric equilibrium structure geometry of benzene is obtained. At other
angles, the distortion introduces alternating shorter and longer carbon-carbon
bonds.  For this system Boguslawski \emph{et al.}~\cite{Boguslawski_2014c}
showed that benzene ($\theta = \ang{60}$) is not the equilibrium structure
within the seniority-zero subspace; an artificial symmetry breaking occurs when
allowing orbital optimization. In this paper we use the experimental geometry
of benzene~\cite{NIST_CCCBDB, Herzberg_1966} and distort the angle while
keeping the atomic distances to the center of mass intact; we did not perform a
geometry optimization while distorting the angle.

We use DOCI-optimized orbitals in the STO-6G basis set to study this artificial
symmetry breaking with all electrons correlated. We chose STO-6G as the
distortion angle of the minimal energy DOCI structure is particularly large for
this basis set. The tensor network calculations are executed with a reduced
bond dimension of 1000.

In \autoref{fig:benzene} the results for the ground state in the different
seniority subspaces are given. In accordance with Boguslawski
\emph{et al.}~\cite{Boguslawski_2014c}, we notice that, indeed, the ground
state is not found at $\theta = \ang{60}$ in the seniority zero subspace. When
the breaking of one pair is allowed in this orbital set, the correction is
rather small and the correct symmetry is not restored; as expected due to the
aforementioned first order decoupling of the seniority-zero and seniority-two
subspace in these orbitals. 

\begin{figure}[!ht]
  \centering
  \includegraphics{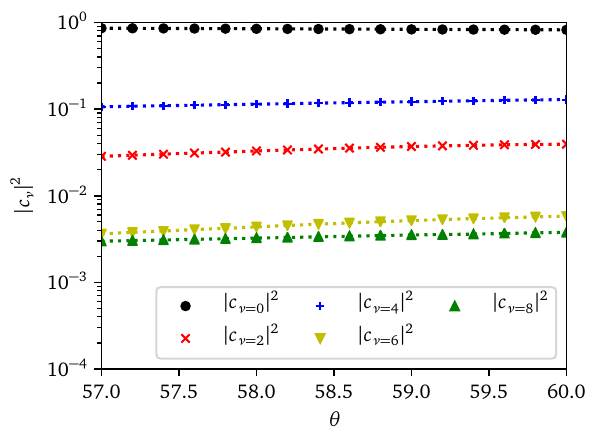}
  \caption{%
    Weights of the different seniority subspaces for the ground state wave
    function of the in-plane distorted benzene in DOCI-optimized orbitals.
  }
  \label{fig:ci_benzene}
\end{figure}

When including progressively higher seniorities, the stable configuration moves
closer to the expected $D_{6h}$ symmetric benzene. The potential energy surface
enjoys a large qualitative correction when including the seniority-four
subspace in the calculations. However the predicted most stable configuration
is still off by $\ang{0.59}$. The inclusion of seniority-six further
improves the quality of the potential energy surface, but only at
seniority-eight the correct symmetry seems to be recovered, at least up to
the resolution of our performed calculations. At this point, the results become
very close to the full seniority results. In \autoref{fig:ci_benzene}, the
weights of the different seniority sectors in the ground state during
distortion are also given. These weights do not express the large changes as
were seen during dissociation of the nitrogen dimer in \autoref{fig:N2_ci}.
This is quite expected as the bond breaking of the nitrogen dimer is a far more
outspoken change than the small benzene distortions in this section.

\subsection{Neon dimer}
\label{sec:Neon}
The neon dimer, constituted by just two noble gas atoms, is very weakly
bound. Although the electrons do not form covalent bonds between the two
atoms, it expresses some bonding character due to weak dispersion forces. In
Ref.~\cite{Aziz_1989}, an empirically fitted potential curve results in a
binding energy of $\SI{-134}{\micro\hartree}$ and a binding distance of
$\SI{3.091}{\angstrom}$.

As the binding of the neon dimer is rather weak and due to dynamical
correlations, it will be very sensitive to the chosen basis set size. For an
accurate description of the potential energy curve, a large basis set should be
chosen and basis set superposition errors (BSSE) should be taken into account
appropriately~\cite{Helgaker_2014_ch8}.
A clear example of the importance of BSSE-corrections is the dissociation curve
on the Hartree-Fock level. At this level of theory no binding is expected as
the Hartree-Fock solution is dispersion-free. However, when using small basis
sets, one would find a binding neon dimer at the Hartree-Fock level if one
neglects to correct the BSSE~\cite{Helgaker_2014_ch8}.

We study the neon dimer in the aug-cc-pVDZ basis; it was found that this basis
set has a favorable tradeoff between mitigating BSSE and numerical stability
issues of larger basis sets. Calculations with different seniority sectors are
executed while using DOCI-optimized orbitals with a frozen $1s$ core. Reduced
bond dimensions up to 800 are used for the DMRG calculations. As the
aug-cc-pVDZ basis is a rather small basis for capturing dispersion forces,
appropriately removing BSSE is important. This is done by using the Boys and
Bernardi counterpoise correction~\cite{Boys_1970}.

\begin{figure}[!ht]
  \centering
  \includegraphics{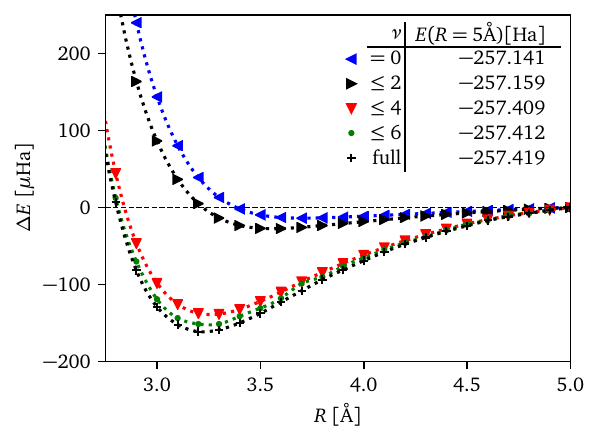}
  \caption{%
    Dissociation curves for the neon dimer without BSSE-correction. The
    energies at large separation distances is given in the inset.
  }
  \label{fig:Ne2_dz}
\end{figure}

In \autoref{fig:Ne2_dz} the raw uncorrected results are given for the
different calculations. For all seniority calculations the neon dimer seems to
be bound. However, for seniority-zero and seniority-two-plus-zero is is 
very weakly bound; only for $\nu \leq 4$ calculations and higher the
neon dimer bounds qualitatively corresponding with the full seniority case.
For the counterpoise correction, equivalent calculations as for the dimer are
executed but where one neon atom is replaced by a chargeless, electronless
ghost atom. This way, we can approximately correct for the extra stabilization
each neon monomer experiences by the extra added basis functions of the other
monomer. The BSSE-corrected dissociation energy for the dimer at distance $r$
is then given by
\begin{equation}
  E_\mathrm{dissoc}(r) = E_\mathrm{Ne-Ne}(r) - E_\mathrm{Ne-ghost}(r) - E_\mathrm{ghost-Ne}(r)~.
\end{equation}

\begin{figure}[!ht]
  \centering
  \includegraphics{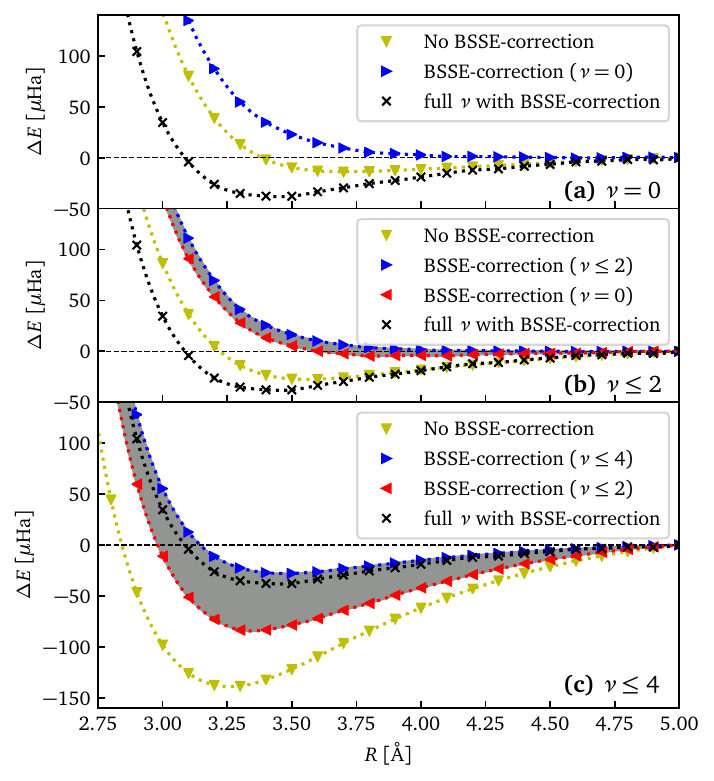}
  \caption{%
    Dissociation curves for the neon dimer with and without BSSE correction for
    different seniority subspaces. When BSSE-correction is performed, the used
    seniority subspace for the ghost atom calculation is given in brackets in
    the legend. The dissociation curve for a full-seniority calculation with
    full BSSE-correction is also shown in each subfigure. Results are shown for
    $\nu = 0$ (a), $\nu \leq 2$ (b) and $\nu \leq 4$ (c) subspace calculations.
  }
  \label{fig:Ne2_dz_BSSE}
\end{figure}

The same level of theory should be used for these ghost calculations
as for the original calculation. This poses a difficulty since the seniority
restricted calculations are not size consistent;
$E_\mathrm{dissoc}(r \to \infty)$ does not tend to zero as is desired.
Assume we have executed a dimer calculation with $\nu \leq 4$, using ghost
calculations with the same $\nu \leq 4$ will over-correct, while ghost
calculations at the lower $\nu \leq 2$ will under-correct. We try to solve this
problem by both over- and under-correcting and shift both curves to 0 in
the dissociation limit. Results for these BSSE-corrections are given for
different seniority sectors in \autoref{fig:Ne2_dz_BSSE} with the grayed area
indicating where the exact BSSE-correction is expected to be.  From
\autoref{fig:Ne2_dz_BSSE}a and \autoref{fig:Ne2_dz_BSSE}b, it seems that the
weak bound present in \autoref{fig:Ne2_dz} for $\nu = 0$ and $\nu
\leq 2$ practically or completely disappears when taking BSSE-corrections into
account. When correcting $\nu \leq 4$ calculations, the over-corrected
dissociation underestimates the dissociation energy a bit with respect to the
BSSE-corrected full-seniority tensor network calculations (which approximate
FCI) while the under-corrected dissociation overestimates the dissociation
energy, as can be expected.

It seems thus that calculations with seniority-zero and seniority-two do not
model the needed dispersion and at least seniority-four is needed. Taking into
account that the DOCI-optimized orbitals are localized on separate Neon atoms
for larger separations, this suggests the breaking of at least one electron
pair at each \ce{Ne} atom is needed, inducing polarization effects in each atom
which give rise to the dispersion energy.   

Finally, we notice that the full-seniority dissociation energy with
BSSE-correction is a factor of three smaller than empirical measurements and
the bond length is overestimated. This is quite normal when studying dynamical
correlations in small basis sets. The limited basis set does not allow all the
needed flexibility for the stabilization of the dimer.

\section{Conclusion}
In this paper, the concept of seniority is joined with tensor network methods.
By using seniority-invariant tensors in a tensor network, we can force all the
renormalized states in the virtual bonds to be eigenstates of the seniority
operator. This allows for arbitrary seniority-restricted calculations. For DOCI
(doubly occupied configuration interaction) calculations, we can immediately
implement the DOCI-projected quantum chemical Hamiltonian in
\autoref{eq:HDOCI}. This results in a very fast tensor network calculation,
partly because of the simpler Hamiltonian, partly because the correlations in
the seniority-zero subspace for molecular systems are easily captured by tensor
networks; even for very large systems, a bond dimension of less than 100
suffices for energies within chemical accuracy of the exact DOCI energy.

The seniority-restricted tensor network method opens up novel ways for
efficient approximate DOCI algorithms with orbital optimization. As one-body
and two-body reduced density matrices are easily extracted from the TNS, one
could alternate between DOCI-TNS calculations and orbital optimizations by
using e.g.\  Newton-Raphson based algorithms~\cite{Roos_1980} or Jacobi
rotations~\cite{Raffenetti_1993, Poelmans_2015b}. As in
Ref.~\cite{Krumnow_2016}, one could also intertwine the orbital optimization
with the tensor network calculation itself. Instead of lowering the
entanglement, one could now optimize for the DOCI energy or perform a seniority
zero-plus-two calculation and minimize the seniority two contribution by
orbital rotations. A proxy for decoupling the seniority-zero and seniority-two
subspaces has been previously done in Ref.~\cite{Boguslawski_2014b}.

Several systems are studied within different seniority subspaces. For the
dissociation of the nitrogen dimer, only a quantitative dissociation curve can
be obtained when at least two pairs are allowed to be broken. This can be
theoretically explained due to Hund's rule. The in-plane distortion of benzene
and its artificial $D_{6h}$ symmetry breaking in the seniority-zero
subspace~\cite{Boguslawski_2014c} is also studied. A large correction of the
artificial symmetry breaking occurs when including seniority-four, however up
to eight unpaired electrons are needed for a complete restoration of the
correct benzene point group symmetry in the used basis set. Finally, also the
dissociation of the neon dimer is considered. At the seniority-zero level of
theory the neon dimer is non-binding; DOCI does not capture the dispersion
forces needed for the weak binding characteristic of neon. Only at
seniority-four and onward, the dispersion forces are adequately picked up.

For all systems, the seniority-two subspace has only a small contribution to
the total wave function when using DOCI-optimized orbitals; as expected by the
theoretical first order decoupling between seniority-zero and seniority-two
subspaces in these types of orbitals~\cite{Boguslawski_2014b}. However, a first
order decoupling is not an exact decoupling and other orbital sets can be found
which attribute even less importance to the seniority-two subspace. An example
of this is given by the natural orbitals of the nitrogen dimer in
\autoref{fig:N2}c.

\section*{Acknowledgements}
KG acknowledges financial support from the Research Foundation Flanders (FWO
Vlaanderen). This research was undertaken, in part, thanks to funding from the
Canada Research Chairs Program (SDB).
Computational resources  and services were provided by Ghent University (Stevin
Supercomputer Infrastructure), Compute Canada, the organization
responsible for digital research infrastructure in Canada, and ACENET, the
regional partner in Atlantic Canada.  ACENET is funded by the Canada Foundation
for Innovation (CFI), and the provinces of New Brunswick, Newfoundland \&
Labrador, and Nova Scotia.
Discussions with Patrick Bultinck are gratefully acknowledged.

\bibliography{mybib.bib}
\nolinenumbers
\end{document}